\documentclass{IEEEtaes}

\usepackage{color}
\usepackage{graphicx}
\usepackage{amsmath,amsfonts}
\usepackage{algorithmic}
\usepackage{algorithm}
\usepackage{array}
\usepackage[caption=false,font=normalsize,labelfont=sf,textfont=sf]{subfig}
\usepackage{textcomp}
\usepackage{stfloats}
\usepackage{url}
\usepackage{verbatim}
\usepackage{cite}
\usepackage{bbm}
\usepackage{xcolor}

\jvol{XX}
\jnum{XX}
\jmonth{XXXXX}
\paper{1234567}
\pubyear{2022}
\doiinfo{TAES.2022.Doi Number}

\newcommand{\bix}{\boldsymbol{x}}

\newcommand{\ba}{\mathbf{a}}

\newcommand{\bd}{\mathbf{d}}

\newcommand{\br}{\mathbf{r}}

\newcommand{\bu}{\mathbf{u}}

\newcommand{\bw}{\mathbf{w}}
\newcommand{\bx}{\mathbf{x}}
\newcommand{\by}{\mathbf{y}}
\newcommand{\bz}{\mathbf{z}}
\newcommand{\bnabla}{\boldsymbol{\nabla}}
\newcommand{\brho}{\boldsymbol{\rho}}
\newcommand{\bgamma}{\boldsymbol{\gamma}}

\newcommand{\bX}{\mathbf{X}}

\newcommand{\bZ}{\mathbf{Z}}
\newcommand{\bbC}{\mathbb{C}}
\newcommand{\bbF}{\mathbb{F}}
\newcommand{\bbR}{\mathbb{R}}

\newcommand{\bbP}{\mathbb{P}}

\newcommand{\bbU}{\mathbb{U}}
\newcommand{\bbmR}{\mathbbm{R}}

\newcommand{\calF}{\mathcal{F}}
\newcommand{\calG}{\mathcal{G}}
\newcommand{\calH}{\mathcal{H}}
\newcommand{\calI}{\mathcal{I}}
\newcommand{\calJ}{\mathcal{J}}
\newcommand{\calK}{\mathcal{K}}

\newcommand{\calP}{\mathcal{P}}

\newcommand{\calR}{\mathcal{R}}
\newcommand{\calT}{\mathcal{T}}
\newcommand{\calZ}{\mathcal{Z}}

\DeclareMathOperator*{\argmin}{argmin}
\DeclareMathOperator*{\argmax}{argmax}

\begin{document}

\title{Interferometric Passive Radar Imaging with Deep Denoising Priors}

\author{Samia Kazemi}
\affil{Rensselaer Polytechnic Institute, Troy, NY 12180, USA} 

\author{Bariscan Yonel}
\member{Member, IEEE}
\affil{Rensselaer Polytechnic Institute, Troy, NY 12180, USA} 

\author{Birsen Yazici}
\member{Fellow, IEEE}
\affil{Rensselaer Polytechnic Institute, Troy, NY 12180, USA}


\receiveddate{Manuscript received October 1, 2022. This work was supported in part by the Air Force Office of Scientific Research (AFOSR) under the agreement FA9550-19-1-0284, in part by Office of Naval Research (ONR) under the agreement N00014-18-1-2068, in part by the National Science Foundation (NSF) under Grant No ECCS-1809234 and in part by the United States Naval Research Laboratory (NRL) under the agreement N00173-21-1-G007.}
\corresp{{\itshape (Corresponding author: B. Yazici)}}

\authoraddress{The authors are with the Department of Electrical, Computer and Systems Engineering, Rensselaer Polytechnic Institute, Troy, NY 12180 USA (e-mail: kazems@rpi.edu; yonelb2@rpi.edu; yazici@ecse.rpi.edu).}


\markboth{AUTHOR ET AL.}{SHORT ARTICLE TITLE}
\maketitle

\begin{abstract}
Passive radar has advantages over its active counterpart in terms of cost and stealth. 
In this paper, we address passive radar imaging problem by interferometric inversion using a spectral estimation method with a priori information within a deep learning (DL) framework.
Cross-correlating the received signals from different look directions mitigates the influence of shared transmitter related phase components despite lack of a cooperative transmitter, and permits tractable inference via interferometric inversion. 
To this end, we leverage deep architectures for modeling a priori information and for improving sample efficiency of state-of-the-art interferometric inversion methods.  
Our approach comprises of an iterative algorithm based on generalizing the power method, and applies denoisers using plug-and-play (PnP) and regularization by denoising (RED) techniques.
We evaluate our approach using simulated data for passive synthetic aperture radar (SAR) by using convolutional neural networks (CNN) as denoisers, and compare our results with state-of-the-art.
The numerical experiment shows that our method can achieve faster reconstruction and superior image quality in sample starved regimes than the state-of-the-art passive interferometric imaging algorithms.
\end{abstract}


\section{INTRODUCTION}
\subsection{Problem Statement}
T{\scshape his} paper studies passive interferometric imaging, which involves the recovery of a signal of interest from the cross-correlations of its linear measurements collected in a spatially diverse sensing geometry.
For such imaging geometries, let $k = 1, \cdots K$ correspond to the frequency samples over the transmission band $\omega \in [\omega_c - B/2, \omega_c + B/2]$ used in the acquisition system, whereas $i, j \in \mathit{S}$ index the location of the receivers, with $|\mathit{S}| = S$. 
Let $\mathbf{a}^k_i, \mathbf{a}^k_j \in \mathbb{C}^N$ denote the sampling vectors corresponding to the $i^{th}$ and $j^{th}$ sensors at a given frequency $\omega_k$, and $\brho^* \in \mathbb{C}^N$ be the ground truth/signal of interest.
Consider the measurement matrix $\mathbf{A}^k$ per frequency, where $\mathbf{a}^k_{i, j}$ are the two distinct columns such that:
\begin{equation}
    \mathbf{f}^k = (\mathbf{A}^k)^H \brho^*, \quad k = 1, \cdots, K
\end{equation}
\begin{equation}
\text{with} \quad f^k_i = \langle \mathbf{a}^k_i , \brho^* \rangle, \quad f^k_j = \langle \mathbf{a}^k_j , \brho^* \rangle,
\end{equation}
as the linear measurements at each receive location. 
The cross-correlated measurements from each location pair $(i, j)$ correspond to the interferometric measurements in frequency domain, as:
\begin{equation}\label{eq:interferom}
d^k_{ij}  = f^k_i  \overline{f^k_j} =  (\mathbf{a}^k_i )^H \brho^* (\brho^*)^H \mathbf{a}^k_j \quad k = 1, \cdots K,
\end{equation}
where $\overline{( \cdot )}$ denotes complex conjugation. Thus, interferometric inversion involves recovery of $\brho^* \in \mathbb{C}^N$  from $d^k_{ij} \in \mathbb{C}, \ k=1,...,K$ under the quadratic model in \eqref{eq:interferom}.

In essence, this is equivalent to recovering $\brho^*$ from the collection of rank-1, data scatter matrices of:
\begin{equation}
\mathbf{D}^k := \mathbf{f}^k (\mathbf{f}^k)^H = \mathbf{A}^k \brho^* (\brho^*)^H (\mathbf{A}^k)^H, \ \ \end{equation}
where  $\mathbf{f}^k = [f^k_1, f^k_2, \cdots f^k_S ]^T$, with \eqref{eq:interferom} corresponding to the upper triangular entries of $\mathbf{D}^k$ per each $k$. 
In this generic form, interferometric inversion problem arises in many applications in different disciplines.
These include radar and sonar interferometry \cite{goldstein1987interferometric, bamler1998synthetic, saebo2010seafloor}, passive imaging in acoustic, electromagnetic and geophysical applications \cite{LWang12_b, Mason2015, son2015passive, Wang13_3, Wang14,  LWang12, wang12, son2017passive, Wacks14, Wacks14_2, ammari2013passive}, and beamforming and sensor localization in large area networks~\cite{patwari2005locating, stoica2007probing} among others. 
In wave-based imaging, correlations were shown to provide robustness to statistical fluctuations in scattering media or incoherent sources \cite{garnier2005imaging, lobkis2001emergence}, and with respect to phase errors in the correlated linear transformations \cite{blomgren2002super, gough2004displaced, mason2016robustness}. 

In this paper, we leverage interferometric inversion for the purpose of addressing the passive radar imaging problem. 
Passive radar systems do not use their own dedicated transmitters, and instead use scattered ambient signals originating from a source of opportunity. 
As a result, passive radar systems are realizable with small mobile receivers that operate with long acquisition modes, providing  spatial diversity and robustness in challenging sensing environments. 
In the setting that illuminators are non-cooperative, precise transmitter location and waveform are unavailable at the receive end to describe the underlying forward mapping for the inversion task.  
To this end, cross-correlating the measurements from different receive locations mitigate the influence of transmitter related terms by removal of the shared phase components.  
Hence, the main motivation for interferometric processing in passive imaging applications is to instead describe a model $\{ \tilde{\mathbf{A}}^k \}_{k = 1}^K$ that can accurately facilitate the inversion in lieu of the ideal but only partially known $\{ {\mathbf{A}}^k \}_{k = 1}^K$, having per $ k = 1, \cdots, K$:
\begin{equation}\label{eq:newinter}
\mathbf{A}^k \brho^* (\brho^*)^H (\mathbf{A}^k)^H \approx \tilde{\mathbf{A}}^k \brho^* (\brho^*)^H (\tilde{\mathbf{A}}^k )^H := \tilde{\mathbf{f}}^k (\tilde{ \mathbf{f}}^k )^H. 
\end{equation}

The key consideration of solving interferometric inversion is in mitigating the partial loss of phase information. 
Beyond the removal of undesirable phase components within the data, the correlation operation results in fundamental limitations in direct factorization of \eqref{eq:newinter}. 
Without access to the full $KS \times KS$ data scatter matrix\footnote{Clearly, one could compute the full scatter matrix as well. However, this would not result in the tractable model used in \eqref{eq:newinter}, thus would not be conductive to the inverse problem at hand.}, i.e., with correlations only computed per fixed frequency, retrieval of the equalized data $\tilde{\mathbf{f}}^k$ by a rank-1 decomposition results in $k$-dependent arbitrary factors of $\mathrm{e}^{j \phi_k}$. This is a consequence of the quadratic nature of the interferometric data matrix via invariance to global phase multipliers. 
As a result, direct factorization demands a crucial phase synchronization step, which increases the number of unknowns to $K + N$ and requires the use of underlying common parameterization 
with respect to the unknown of interest $\brho^*$. 
The performance of such formulation then strongly hinges on the accurate recovery of the phase factors. 
This is undesirable as small phase errors are known to yield drastic errors in the reconstructed imagery \cite{mason2016robustness}. 

Ultimately, using the underlying parameterization of the scene is necessary for the feasibility of the resulting interferometric inversion problem. 
This motivates approaches for direct inversion from the quadratic measurement model of \eqref{eq:interferom} to avoid inducing phase ambiguity over frequency samples, which form the state-of-the art. 

\subsection{Prior Art and Motivation}

Conventionally, interferometric inversion in imaging applications has been approached by Fourier based techniques, such as time or frequency difference of arrival (TDOA/FDOA) backprojection \cite{Yarman10, Wang11, wang12, Wang13_3, wacks2018doppler, Wacks14, Wacks14_2}.
While these methods are practical and computationally efficient, their applicability is limited to scenes composed of well-separated point targets due to underlying assumptions.
As an alternative, low rank matrix recovery (LRMR) theory has been explored for interferometric inversion \cite{Mason2015, Demanet13}.
Notably, these solvers are inspired by the PhaseLift method \cite{Candes13a, Candes13b, Demanet14}, hence suffer from the same drawbacks in computation and memory to semi-definite programming (SDP) in practice. 
In \cite{Mason2015}, an iterative optimization approach to LRMR was developed for interferometric passive imaging to circumvent the poor scaling properties of SDP approaches.
While this method is more efficient than the SDP solvers, still operates by squaring the number of unknowns, hence still requires significant memory and computational resources for imaging. 
Additionally, these convexified lifting based solvers require stringent theoretical conditions on the measurement model, 
which poses a major theoretical barrier for interferometric inversion problems with deterministic forward models. 

Motivated by the reduced computational complexity and memory requirements of non-convex optimization over the lifting based methods in phase retrieval literature \cite{candes2015phase_IEEE}, we developed the generalized Wirtinger Flow (GWF) for interferometric inversion in \cite{yonel2019generalization}.
Namely, GWF provides deterministic exact recovery guarantees to a general class of problems that are characterized over the equivalent lifted domain by the restricted isometry property (RIP) on the set of rank-1, positive semi-definite (PSD) matrices, while operating solely on the original signal domain.
In \cite{yonelTCI}, we established the sufficient condition of exact recovery for passive imaging on multi-static geometry, where we determined the physical parameters of the system to ensure exact recovery. 
Furthermore, we introduced theoretical framework that facilitated a resolution analysis and tractable sample complexity of interferometric wave-based imaging under the far-field and small scene assumptions, and showed that the GWF algorithm achieves super-resolution in parameter regimes that commonly correspond to passive settings using $O(\sqrt{N})$ distinct look directions.

Despite its impact in theoretical outcomes, the GWF approach has certain limitations. 
The data-rates from correlations grow with $S^2$ in number of look directions, which amounts to $\mathcal{O}(N^{3/2})$ total sample complexity in \cite{yonelTCI} for super-resolution capability, and $\mathcal{O}(N^{5/4})$ for minimal feasibility. 
Recently, a distributed analogoue of GWF was developed in \cite{farrell2022distributed} with tunable graph connectivity in forming local subset of correlations within the sensor geometry, hence provides control on growth of the data-rates for inference without sacrificing performance guarantees. 
Still, even with $\mathcal{O}(N^{5/4})$ minimal sample complexity, there exists an $\sqrt{L}$-factor growth in establishing the sufficient condition, which indicates break-down of GWF guarantees in imaging scenes beyond a critical length. 
Other limitations include oversampling requirements on $K$, poor scaling of the sufficient condition bounds with respect to the imaging aperture/field of view, and slow convergence due to the first-order nature of the algorithm updates.  

Our motivation in this paper is to address the shortcomings of GWF by leveraging a priori information in the form of a constrained spectral estimation approach. 
Spectral methods form the crucial initial step of non-convex phase retrieval techniques which uses the spectra of the back-projection estimate in the lifted domain \cite{yonel2022spectral}.  
Incorporating constraints that capture image features offers to morph the signal space during the spectral search and yield structurally sound estimates without the need of iterative updates in GWF.
As a result, structural prior information provides the potential to
improve the computational and acquisition efficiency of interferometric passive radar imaging by decreasing the order of required number of look directions or the oversampling factors in frequency \cite{akhtar2021passive}. 
To this end, we utilize deep architectures to learn effective representations of the signal manifold.



\subsection{Our Approach and Contribution}
Our objective in this paper is to apply deep learning (DL) for designing an interferometric imaging algorithm using a priori information. 
We are particularly interested in deep models 
to capture structure in the underlying scene with approximation capability beyond that is achievable by denoising with sparsity-based functional regularizer. 
For our approach, we consider regularization within the spectral estimation framework, indirectly and directly, by applying a denoising operator using plug-and-play (PnP) \cite{venkatakrishnan2013plug, kamilov2017plug_nonlinear, sun2019PnP} and regularization by denoising (RED) \cite{romano2017little, Metzler2018_prdeep} approaches within DL frameworks, respectively.

The spectral matrix estimates in practically relevant interferometric imaging geometries may not possess the favorable characteristics such that its leading eigenvector retains important structural information about the unknown. 
This is especially the case when operating below the requirements identified for feasibility of the GWF theory, where the standard spectral method is not guaranteed to preserve sufficient similarity on the underlying scene of interest. 
Significant difficulties in denoiser training are posed in such problem settings, as it is challenging to determine the statistics for noisy training images that enables reconstruction within a few iterations to limit the computation cost, which forms the basis of PnP and RED frameworks.
As a solution, we implement our denoiser based algorithms by using the unrolling technique with the network depth increased sequentially at each training instance.
Aside from easier denoiser training, unrolling has the added benefit of reducing training data requirement, which is particularly desirable for SAR to minimize operational costs.

We implement our imaging networks for simulated passive SAR dataset by using convolutional neural networks (CNN) architectures for denoising, and compare performances to that of the spectral initialization approach, applied in the GWF algorithm~\cite{yonel2019generalization}.
Furthermore, we consider its variant using sparsity prior, as well as other state-of-the-art interferometric imaging techniques, and numerically observe the expected benefits in reduced sample complexity and faster convergence. 

\subsection{Organization}
Rest of this paper is organized as follows: In Section II, we present our received signal model associated with the interferometric imaging task for passive bistatic SAR.
In Section III, we present relevant background information, introduce our denoising prior and spectral estimation based imaging algorithms and discuss their DL-based implementation details.
Section IV describes our observations from a range of numerical experimentation on simulated SAR datasets.
Finally, Section V concludes our paper.

\section{Received Signal Model}
\label{sec:problem_statement}
\begin{figure}
\centering
\includegraphics[width=0.7\columnwidth]{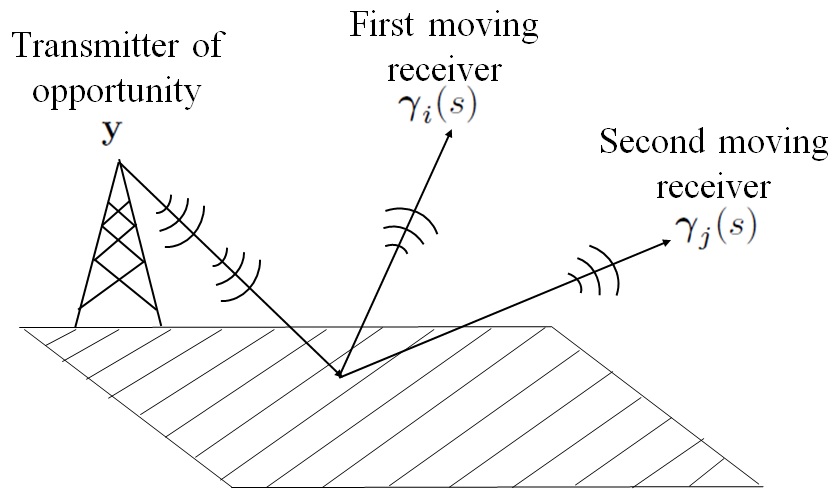}
\caption{Passive bistatic SAR geometry.}
\label{fig:problem_statement_diagram}
\end{figure}
We consider the passive bi-static SAR imaging configuration with a single stationary transmitter, and two moving airborne receivers, whose trajectories are spatially separated throughout the data-collection process\footnote{We note that our interferometric imaging approach similarly applies to the case with multiple stationary receivers.}.
This passive SAR imaging configuration is shown in Fig.~\ref{fig:problem_statement_diagram}.
We assume that the area being imaged has a flat topography, and the locations within the scene are characterized by $\bx = [\bix, 0]\in\bbR^3$, where $\bix\in\bbR^2$ indicates a 2-D location in ground plane.
Let the stationary transmitter be located at $\by\in\bbR^3$, and let $\rho:\bbR^2\mapsto\bbmR$ be the ground reflectivity function.
Suppose the frequency samples of the reflected wave are collected at $S$ slow-time points by the two receivers.
We represent the locations of the two receivers at slow time $s\in[S]$ by $\bgamma_i(s)\in\bbR^3$, $i = 1, 2$.
For all $s\in[S]$, the trajectories of the airborne receivers are such that $\bgamma_i(s) \neq \bgamma_j(s)$.
We represent the speed of transmission propagation through the background medium and the fast-time frequency by $c_0\in\bbmR^+$ and $\omega\in\bbmR^+$, respectively.
Let $\omega\in[\omega_c - \frac{B}{2}, \omega_c + \frac{B}{2}]$ denote the fast-time frequency with $\omega_c$ being the center frequency.

Under the Born and start-stop approximations, the received signal model is~\cite{Mason2015}
\begin{align}
     \label{eq:received_signal_a} f_i(\omega, s) & = \int e^{i\frac{\omega}{c_0}\phi_i(\bx, \by, s)}A_i(\bx, \omega, s)\rho(\bix)d\bix,
\end{align}
where
\begin{align}\label{eq:phaseDelay}
    \phi_i(\bx, \by, s) & = |\bx - \bgamma_i(s)| + |\bx - \by|.
\end{align}
The amplitude term $A_i$ is determined by the antenna beampatters and geometric spreading factors.
The cross-correlated measurements from the two receivers evaluated at each slow-time is given as
\begin{align}
    d(\omega, s) & = f_i(\omega, s)\overline{f_j(\omega, s)}.
\end{align}
Using~\eqref{eq:received_signal_a}, and under small-scene and far-field assumptions, the cross-correlated measurements can be modelled as~\cite{Mason2015}:
\begin{align}
     & d(\omega, s) \nonumber \\
     \label{eq:cross_corr_cont} & = \int e^{i\frac{\omega}{c_0}\phi_{ij}(\bx, \bx', \by, s)}A_{ij}(\bx, \bx', \omega, s)\rho(\bix)\bar{\rho}(\bix')d\bix d\bix',
\end{align}
where
\begin{align}
    \phi_{ij}(\bx, \bx', \by, s) & = \tilde{\phi}_i(\bx, \by, s) - \tilde{\phi}_j(\bx', \by, s),
\end{align}
and
\begin{align}
    \label{eq:tilde_phi_i} \tilde{\phi}_i(\bx, \by, s) & = |\bx - \bgamma_i(s)| + \hat{\by}.\bx,
\end{align}
with $\hat{\by}$ being the unit vector in the direction of $\by$.
$A_{ij}(\bx, \bx', \omega, s)$ relates to the antenna beam-patterns, $J_i(\bx, \omega)$, $J_j(\bx', \omega)$ and $J_t(\bx, \omega)$, as
\begin{align}
A_{ij}(\bx, \bx', \omega, s) & \approx \frac{J_i(\bx, \omega)\bar{J_j}(\bx', \omega)C^2_t}{|\bgamma_i(s)||\bgamma_j(s)||\by|^2},
\end{align}
where $|J_t(\bx, \omega)| \approx C_t \in \mathbbm{R}^+$, under the assumption that the transmitted waveform has a flat spectrum, and $-3$dB beam-width encompasses the area being imaged~\cite{Mason2015, yonelTCI, akhtar2021passive}.

Our objective is to recover $\rho$ using the model~\eqref{eq:cross_corr_cont} and a priori information on $\rho$.
Towards this objective, we proceed by first discretizing the scene into $N$ points, with locations denoted by $\{\bix_n\}_{n = 1}^N$ where $\bix_n\in\bbR^2$, and define a corresponding ground truth image vector $\brho^*\in\bbC^N$ as
\begin{align}
    \brho^* = \begin{bmatrix}\rho(\bix_1) & \cdots & \rho(\bix_N)\end{bmatrix}^T.
\end{align}
Similarly, we consider $K$ discrete fast-time frequency samples, $\{\omega_k\}_{k = 1}^K$, sampled uniformly within the band $[\omega_c - B / 2, \omega_c + B / 2]$ to form a discretized data vector $\bd\in\bbC^M$ with $M$ representing the total number of measurements, i.e., $M = SK$.

Cross-correlated measurement $d(\omega_k, s)$ can be represented under this modified data model using linear sampling vector, $\ba^{k, s}_i\in\bbC^N$ for $i = 1, 2$, $k\in[K]$ and $s\in[S]$, as
\begin{align}
    d(\omega_k, s) & = \langle\ba^{k, s}_i, \brho^*\rangle\overline{\langle\ba^{k, s}_j, \brho^*\rangle},
\end{align}
where
\begin{multline*}
\ba^{k, s}_i = 
[\begin{matrix} e^{i\frac{\omega_k}{c_0}\tilde{\phi}_i(\bx_1, \by, s)}A_i(\bx_1, \omega_k, s) & \cdots \end{matrix} \\
 \begin{matrix} e^{i\frac{\omega_k}{c_0}\tilde{\phi}_i(\bx_N, \by, s)}A_i(\bx_N, \omega_k, s) \end{matrix}]^H,
\end{multline*}
with the terms from \eqref{eq:tilde_phi_i} and
\begin{align}
    A_i(\bx_n, \omega_k, s) & = \frac{J_i(\bx_n, \omega)C_t}{|\bgamma_i(s)||\by|},
\end{align}
for $n\in[N]$.
$\bd\in\bbC^M$ relates to $d(\omega, s)$ as
\begin{align}
    & \bd = \nonumber \\
    & \begin{bmatrix}
    d(\omega_1, 1) & \ldots & d(\omega_1, S) & d(\omega_2, 1) & \ldots & d(\omega_K, S)
    \end{bmatrix}^T.
\end{align}
Let $\calF:\bbC^{N \times N} \longrightarrow \bbC^M$ be a linear lifted forward mapping operator defined such that,
\begin{align}
    \label{eq:data_model} \bd & = \calF(\brho^*{\brho^*}^H).
\end{align}
Our aim to estimate $\brho^*$ directly from the known cross-correlated measurement related vector, $d$, and the fully-known imaging geometry related operator $\calF$, by using structural prior information about the image class of interest.

\section{Denoising Prior and Spectral Estimation-based Interferometric Imaging Network}
\subsection{Background on Methodology}
GWF for interferometric inversion is inspired by the non-convex phase retrieval algorithm in~\cite{candes2015phase_IEEE, yonel2020deterministic}. GWF uses a two-step algorithmic approach to solve quadratic equations involving first a spectral initialization \cite{netrapalli2013phase}, then a simple first-order iterative refinement as follows:
\begin{align}
    \brho_l & = \brho_{l - 1} - \frac{\mu_l}{\|\brho_0\|^2}\bnabla\calJ(\brho)|_{\brho = \brho_{l - 1}}.
\end{align}
$\calJ(\brho)$ is the quadratic objective function associated with the interferometric inversion problem given by
\begin{align}
    \label{eq:objective_function_P} \calJ(\brho) & = \frac{1}{2M}\sum_{k, s = 1}^{K, S}\left[(\ba^{k, s}_i)^H\brho\brho^H\ba^{k, s}_j - d(\omega_k, s)\right]^2,
\end{align}
for the passive SAR problem described in Section \ref{sec:problem_statement}.
The key observation of \cite{yonel2019generalization} is that one can guarantee sufficient accuracy of the initial \emph{spectral estimate}, such that the simple iterations converge to the true solution if the linear forward map in \eqref{eq:data_model} satisfies the restricted isometry property over the set of rank-1, positive semi-definite (PSD) matrices with a restricted isometry constant (RIC)-$\delta \leq 0.214$. 

The spectral initialization step involves setting the leading eigenvector of the following matrix
$\hat{\bX}$ as an initial estimate of $\brho$:
\begin{align}
    \label{eq:GWF_spectral_matrix} \hat{\bX} & := \calP_s(\calF^H(\bd)),
\end{align}
where $\calP_S(\bZ) := \frac{1}{2}(\bZ + \bZ^H)$ for $\bZ\in\bbC^{N \times N}$ and
\begin{align}
    \calF^H(\bd) & = \frac{1}{M}\sum_{k, s = 1}^{K, S}d(\omega_k, s)\ba^{k, s}_i(\ba^{k, s}_j)^H.
\end{align}
Hence, the estimate is obtained by back-projection of data on the lifted domain.

The main premise of this algorithmic framework is in the well-conditioning of the normal operator $\mathcal{F}^H \mathcal{F}$ over terms of the form $\brho \brho^H$, which is controlled by the RIC-$\delta$ bound used in the sufficient condition of GWF. 
However, this has stringent implications on the imaging geometry and the required sample complexities for the validity of theoretical arguments. 
Namely, the interferometric inversion by GWF becomes ill-posed at low ratios of the number of measurements to the number of unknowns, which is quite common for high dimensional imaging problems. 

One way to circumvent these shortcomings while maintaining low computational cost is to incorporate prior information about the unknown image class to the spectral estimation process.
This is due to loss of information at the initialization stage since the normal operator, $\calF^H\calF$, does not approximate an identity map over rank-1, PSD matrices for finite number of measurements. 
We therefore design our approach to utilize prior information about the underlying scene of interest for attaining structurally sound estimates directly via the spectral method. 
As a consequence, we bypass the objective function based optimization criteria of GWF, and consider the penalty for extracting the leading eigenvector of $\hat{\bX}$ by reformulating the power method \cite{journee2010generalized, yuan2013truncated} with regularization. 


\subsection{Spectral Method with Prior Information}

Prior information is commonly incorporated while solving an optimization problem by adding a suitable \emph{regularization} term, $\calR(.):\bbC^N\mapsto\bbmR$, to a \emph{data fidelity} measure $\calK(.):\bbC^N\mapsto\bbmR$ associated with the underlying data model.
This leads to a modified objective function to be minimized to estimate the unknown quantity with $\calR(\brho)$ imposing some structural prior information during the reconstruction process.
For example, under the widely deployed sparsity prior, $\calR(\brho)$ is commonly set as $\eta\|\brho\|_0$ or $\eta\|\brho\|_1$ with $\eta\in\bbmR^+$ denoting an appropriately chosen regularization constant.
In~\cite{akhtar2021passive}, we utilized a truncated power method~\cite{yuan2013truncated} for the initialization phase of an interferometric imaging algorithm to generate an initial image under the assumption that the sparsity level $k$ is known in advance.
In general, sparse leading eigenvector estimation and the sparse PCA are well-studied problems in the literature.

In this paper, we first cast this problem as a minimization task such that the power method arises as a natural consequence of applying a proximal gradient descent (PGD) algorithm when the regularization term is not included during solution.
With the regularization term included, the associated PGD algorithm becomes a realization of the power method with prior.
More specifically, following optimization problem forms the basis of our DL based imaging approach:
\begin{align}
    \label{eq:proximal_power_obj} \brho^* & = \argmin_{\brho\in\bbC^N} \calJ_S(\brho) + \calI_n(\brho) + \calR(\brho).
\end{align}
The data fidelity term $\calJ_S:\bbC^N\mapsto\mathbbm{R}$ is defined as
\begin{align}
    \label{eq:g1_def} \calJ_S(\brho) & = \frac{1}{\gamma}\left(- \beta\brho^H\hat{\bX}\brho + \brho^H\brho\right),
\end{align}
where $\gamma, \beta\in\mathbbm{R}^+$.
$\calI_n:\bbC^N\mapsto\mathbbm{R}^+$ is an indicator function whose output is $0$ when the corresponding input vector has unit $\ell_2$ norm, and it takes large values otherwise.
With an explicitly defined regularization term, the proximal algorithm addressing~\eqref{eq:proximal_power_obj} involves the following set of updates at the $l^{th}$ iteration:
\begin{align}
    \label{eq:pnp_pgd_pm1} \bw_l & = \brho_{l - 1} - \gamma\bnabla\calJ_S(\brho)|_{\brho = \brho_{l - 1}}, \\
    \label{eq:projection_minimization} \by_l & = \argmin_{\bx \in \bbC^N} \|\bx - \bw_l\|^2 + \eta\calR(\bx), \\
    \label{eq:power_method_norm} \brho_l & = \by_l/\|\by_l\| = \calG(\by_l),
\end{align}
where
\begin{align}
    \label{eq:nabla_g1_def} \nabla\calJ_S(\brho) & = (-\beta\hat{\bX}\brho + \brho) / \gamma.
\end{align}
We observe that the update step in~\eqref{eq:pnp_pgd_pm1} simplifies to
\begin{align}
    \by_l & = \beta\hat{\bX}\brho_{l - 1},
\end{align}
which is similar to the step applied during the power method updates.
Note that the normalization step in~\eqref{eq:power_method_norm} is included to account for the indicator function.
The proximal operator in~\eqref{eq:projection_minimization} modifies $\bw_l$ to a neighboring point that satisfies the structural information captured by $\calR$.

However, due to the difficulty in formulating an appropriate $\calR(.)$ in the absence of explicit prior information, and the challenges in designing an $\calR$ that leads to preferably a closed form solution of~\eqref{eq:projection_minimization},we instead design our imaging approach following the PnP and RED frameworks.
For PnP, we know that $\calR(.)$ is not required to be explicitly defined.
Instead, under the assumption that the residual noise after the update step in~\eqref{eq:pnp_pgd_pm1} have i.i.d. Gaussian distribution, the proximal operator in~\eqref{eq:projection_minimization} can be interpreted as a denoiser for a given $\calR$.
Therefore, we can readily design our desired denoising prior-based power method for interferometric imaging under a PnP framework by replacing the minimization problem in~\eqref{eq:projection_minimization} by the following step:
\begin{align}
    \label{eq:pnp_pgd_pm2} \bz_l & = \calZ_0(\bw_l).
\end{align}
where $\calZ_0:\bbC^N\mapsto\bbC^N$ denotes an appropriately designed non-linear operator.
Its output is used to calculate the image estimate $\brho_l$ by using the normalizing operator $\calG$, i.e.,
\begin{align}
    \label{eq:pnp_pgd_pm3} \brho_l & = \bz_l/\|\bz_l\| = \calG(\bz_l).
\end{align}
This PnP based formulation of our power method for interferometric imaging, presented in~\eqref{eq:pnp_pgd_pm1},~\eqref{eq:pnp_pgd_pm2} and~\eqref{eq:pnp_pgd_pm3}, can be represented in a single step as follows:
\begin{align}
    \label{eq:pnp_pgd_combined} \brho_l & = \calT(\brho_{l - 1}) = \calG \circ \calZ_0\left(\beta\hat{\bX}\brho_{l - 1}\right),
\end{align}
where the combined operator $\calT$ is defined as
\begin{align}
    \calT = \calG \circ \calZ_0 \circ (\calI - \gamma\nabla\calJ_S).
\end{align}
\begin{figure}
\centering
\includegraphics[width=0.45\columnwidth]{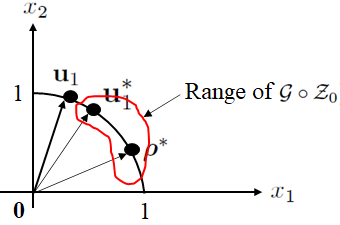}
\caption{Schematic diagram showing $\bu_1$, $\bu^*_1$ and $\brho^*$ for $N = 2$.}
\label{fig_two_D_points}
\end{figure}
We note that $\calZ_0$ captures structural information about the unknown images similar to $\calR$.
This algorithm essentially attempts to recover a point $\bu^*_1\in\bbC^N$ that is located within a small neighborhood of $\bu_1$ such that $\bu^*_1\in\text{Range}(\calG\circ\calZ_0)$.
Here, $\bu^*_1$ can be defined as the best estimation of $\bu_1$ that possesses the desired structural properties encapsulated by $\calZ_0$, i.e.,
\begin{align}
    \bu^*_1 = \argmax_{\bu\in\text{Range}(\calG\circ\calZ_0)}\bu^H\hat{\bX}\bu,
\end{align}
and $(\bu^*_1)^H\hat{\bX}\bu^*_1 \leq \bu^H_1\hat{\bX}\bu_1$.
For the case where $N = 2$, we show a visualization of $\bu_1$, $\bu^*_1$ and $\brho^*$, with $\|\brho^*\| = 1$, in Fig.~\ref{fig_two_D_points}.
This change in the ground truth quantity from $\brho^*$ to $\bu^*_1$ arises from our spectral estimation based formulation, and it reveals the following important desired denoiser property: for all $\brho^*$ from the image class of interest, exact recovery requires $\calZ_0$ to be adequately \emph{precise} for the corresponding $\bu^*_1$ vector to align as closely to $\brho^*$ as possible.
Moreover, this further stems from the reality that for an arbitrary $\calZ_0$, it is difficult in general to explicitly define a corresponding regularization term under which, the PnP algorithm presented in~\eqref{eq:pnp_pgd_combined} achieves the same minimum point as the one attained by the algorithm described in~\eqref{eq:pnp_pgd_pm1} to~\eqref{eq:power_method_norm}.
As a consequence, $\brho^*$ is not necessarily a minimum point of an underlying objective function anymore.
Instead, we are interested in the convergence of our algorithm to a set of fixed points of the combined operator $\calT$.
Let this set be denoted by $\bbF$, i.e. $\bbF = \{\brho\in\bbC^N:\brho = \calT(\brho)\}$.
On the other hand, we represent the global solution set of the unconstrained interferometric inversion problem by $\bbP$, i.e., $\bbP = \{e^{i\phi}\brho^* : \phi \in [0, 2\pi]\}$.
The exact recovery for the interferometric imaging problem using our proposed approach in~\eqref{eq:pnp_pgd_combined} therefore amounts to achieving optimal conditions on the denoiser for the given data fidelity term $\calJ_S$ in~\eqref{eq:g1_def}, such that the iterative updates in~\eqref{eq:pnp_pgd_combined} converge to an element of $\bbF\cap\bbP$.

On the other hand, under the RED framework~\cite{romano2017little, Metzler2018_prdeep}, the regularization term $\calR(\brho)$ is defined explicitly as a function of the denoiser as follows:
\begin{align}
    \label{eq:regularization_RED} \calR(\brho) & = 0.5\brho^H(\brho - \calZ_0(\brho)).
\end{align}
Therefore, we can alternatively modify the power method for interferometric inversion problem by retaining the update steps from~\eqref{eq:pnp_pgd_pm1} and~\eqref{eq:power_method_norm} unchanged and by using the expression of $\calR(.)$ from~\eqref{eq:regularization_RED} in~\eqref{eq:projection_minimization}.
Under the two conditions on the denoiser defined in~\cite{romano2017little}, namely, local homogeneity and strong passivity, it is shown that the solution to the minimization problem for the corresponding proximal operator,  i.e.,
\begin{align}
    \label{eq:projection_minimization_RED} \by^{RED}_{l} & = \argmin_{\bx \in \bbC^N} \|\bx - \bw_l\|^2 + \frac{\eta}{2}\bw^T_l(\bw_l - \calZ_0(\bw_l)),
\end{align}
can be approximated as $\br_{\infty}$, where $\br_j$ for $j\in\{1, 2, ..., \infty\}$ is calculated as
\begin{align}
    \label{eq:red_update} \br_j = \left(\br_{j - 1} + \eta\calZ_0(\br_{j - 1})\right) / (1 + \eta),
\end{align}
and $\br_0$ is set equal to $\bw_l$.

We note that the algorithm described in~\eqref{eq:pnp_pgd_combined} is similar to the projected power method presented in~\cite{liu2022generative}.
However,~\cite{liu2022generative} implements a projection operator $\calP$ instead of the denoising and the normalization step presented in~\eqref{eq:pnp_pgd_pm2} and~\eqref{eq:pnp_pgd_pm3}, respectively.
$\calP$ is defined as $\calP(\bz) = \argmin_{\bw\in\text{Range}(\calH)}\|\bw - \bz\|^2$ with $\calH$ being a pre-trained variational auto-encoder whose range constitutes a subset of the unit sphere.
This optimization problem is addressed by iterative algorithms in~\cite{liu2022generative}.
For example, the Adam optimizer with $200$ updates and a learning rate of $0.03$ was implemented during the numerical simulations in~\cite{liu2022generative}.
On the other hand, for our algorithm in~\eqref{eq:pnp_pgd_combined}, $\calZ_0$ can be interpreted to be modelling a proximal operator for an unknown regularization term $\calR(.)$ such that $\calZ_0(\bz) = \argmin_{\bw\in\bbC^N}\|\bw - \bz\|^2 + \calR(\bw)$.
Unlike~\cite{liu2022generative},
our approach in~\eqref{eq:pnp_pgd_combined} does not explicitly define an associated regularization term $\calR(.)$ for $\calZ_0$, and does not apply any iterative algorithm for denoising.
Our RED based formulation, on the other hand, implements the denoiser within the particular definition of $\calR$ from~\eqref{eq:regularization_RED}.

\subsection{Deep Imaging Network}
\begin{figure*}[!t]
\centering
\subfloat[]{\includegraphics[width=1.1\columnwidth]{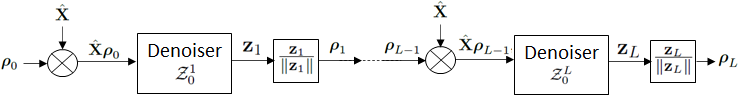}%
\label{fig_schematic_diagram_pnp}}
\\
\subfloat[]{\includegraphics[width=1.2\columnwidth]{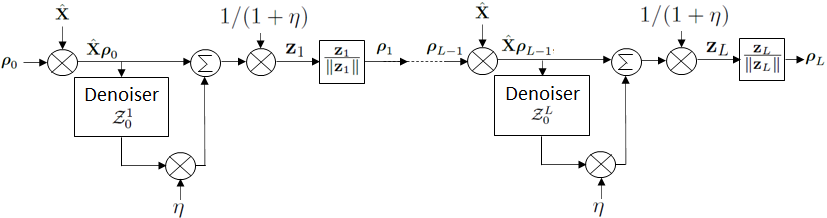}%
\label{fig_schematic_diagram_red}}
\caption{Schematic diagram of imaging network designed based on (a) PnP and (b) RED algorithms.}
\label{fig:schematic_diagram}
\end{figure*}
We begin by noting that the two versions of our denoising prior based imaging algorithms can be implemented with or without applying DL.
However, in this paper, we are aiming to introduce algorithms that perform interferometric inversion under geometries for which, the sufficient conditions for exact recovery of the state-of-the-art algorithms are not necessarily satisfied.
In these challenging regimes, using DL can be particularly beneficial for overcoming the lack of redundancies in the measurements, as well as for relaying the computational cost of incorporating complex prior information to the learning stage instead of the test phase when new measurement samples are used for imaging.

Hence, we utilize DL to implement our denoising prior based algorithms, presented in~\eqref{eq:pnp_pgd_combined} and in~\eqref{eq:pnp_pgd_pm1},~\eqref{eq:projection_minimization_RED} and~\eqref{eq:power_method_norm}, in two stages.
First, similar to the state-of-the-art PnP and RED algorithms, we adopt DNs to model the denoisers instead of using any pre-defined non-linear function for this purpose.
Second, to render our algorithms suitable for the imaging configuration described in Section~\ref{sec:problem_statement}, with physical parameter values such that $\bu_1$ significantly deviates from $\brho^*$, we adopt the unrolling technique~\cite{kazemi2022unrolled} that leads to end-to-end imaging networks.
Unlike PnP and RED algorithms, we apply this technique instead of incorporating arbitrary pre-trained denoisers at the steps of the iterative algorithms implemented using specific stopping criteria.
Separate denoiser training commonly proceeds using a set of noisy images generated using additive Gaussian distributed noise with the clean ground truth images.
Several versions of the denoiser is typically trained using datasets with different noise variances.
However, it is difficult in general to optimally adjust the noise levels, for which the different versions of the denoisers applied at specific iterations are trained, such that the average number of updates necessary for convergence to the fixed points are as small as possible.

Furthermore, our proposed algorithms
apply initial image vectors that are not structured or derived using any sophisticated model-based formulation.
Instead, we utilize either random initialization or some pre-defined fixed normalized vector as the initial point in order to reduce the associated computation cost.
Depending on the choice of the initial vector, its mapping via $\hat{\bX}$ either may not possess the desired i.i.d. Gaussian distribution noise property or its variance may be too large such that a denoiser of a particular capacity cannot remove it sufficiently to recover any useful structural information.
In general, we can infer that if the leading eigenvector $\bu_1$ of $\hat{\bX}$ significantly deviates from $\brho^*$, then $\hat{\bX}\brho^*$ may deviate from $\brho^*$ significantly as well, and hence may not retain much of the useful structural information present in $\brho^*$.
Furthermore, for a particular initial image and the denoiser architecture, the number of iterations required for satisfying the stopping criteria for fixed point detection can be large leading to high computation costs.
Therefore, PnP and RED type realizations of our algorithms are implemented using the unrolling technique to accommodate limited denoising network capacities while keeping the required number of update stages as small as possible.

Our end-to-end imaging networks corresponding to the PnP and RED based algorithms are shown in Fig.~\ref{fig_schematic_diagram_pnp} and~\ref{fig_schematic_diagram_red}, respectively, where the later implements a single step of~\eqref{eq:red_update}.
The initial image $\brho_0$ has unit $\ell_2$ norm.
The set of denoisers, $\{\calZ^i_0\}_{i = 1}^L$, share the same DN architecture with the same set of trainable parameters.
However, these parameter values are learned independently at the various update stages with the goal of attaining an optimal set of denoisers.
Our unrolled imaging network is trained by minimizing a loss function $c_{tr}(\bbU)$, where $\bbU$ denotes the set of trainable parameters of the denoisers and $c_{tr}(\bbU)$ is calculated as $\frac{1}{T}\sum_{t = 1}^T\|\brho^*_t - \brho_{L, t}\|^2$.
Here, $T$ denotes the number of training samples, and $\brho^*_t$ and $\brho_{L, t}$ refers to the $t^{th}$ ground truth image and the output vector, respectively, of the imaging networks in Fig.~\ref{fig:schematic_diagram}.
We gradually increase the number of update stages $L$ during training to utilize as few updates as possible during imaging.


\section{Numerical Simulations}
In this section, we numerically demonstrate the feasibility and performance of our deep denoising prior based imaging networks, presented in Fig.~\ref{fig:schematic_diagram}, using two simulated passive bistatic SAR datasets.
For quantitatively assessing the estimated image quality, we use the normalized mean squared error (MSE) as the figure of merit.
It is calculated as $\mathrm{MSE} = \frac{1}{T_s}\sum_{t = 1}^{T_s}\|\hat{\brho}_t - \brho^*_t\|^2 / \|\brho^*_t\|^2$, where $\hat{\brho}_t$ refers to an estimated image for the $t^{th}$ test sample and $\brho^*_t$ denotes the corresponding ground truth image.
Aside from feasibility verification, we have the following two major objectives for our numerical simulations:
\begin{enumerate}
    \item Compare the performances of our PnP and RED based networks to the ones obtained using the state-of-the-art GWF algorithm as well as the power method and truncated power method generated leading eigenvector estimations, where the later imposes sparsity constraint.
    \item Numerically assess the performance of our imaging networks in the presence of additive noise in the cross-correlated measurements, and empirically validate the important intuition that the deep denoising prior enables improved sample complexity and computation time compared to the state-of-the-art GWF algorithm.
\end{enumerate}

\subsection{Dataset Description}
\begin{figure}[t]
\centering
\includegraphics[width=0.9\columnwidth]{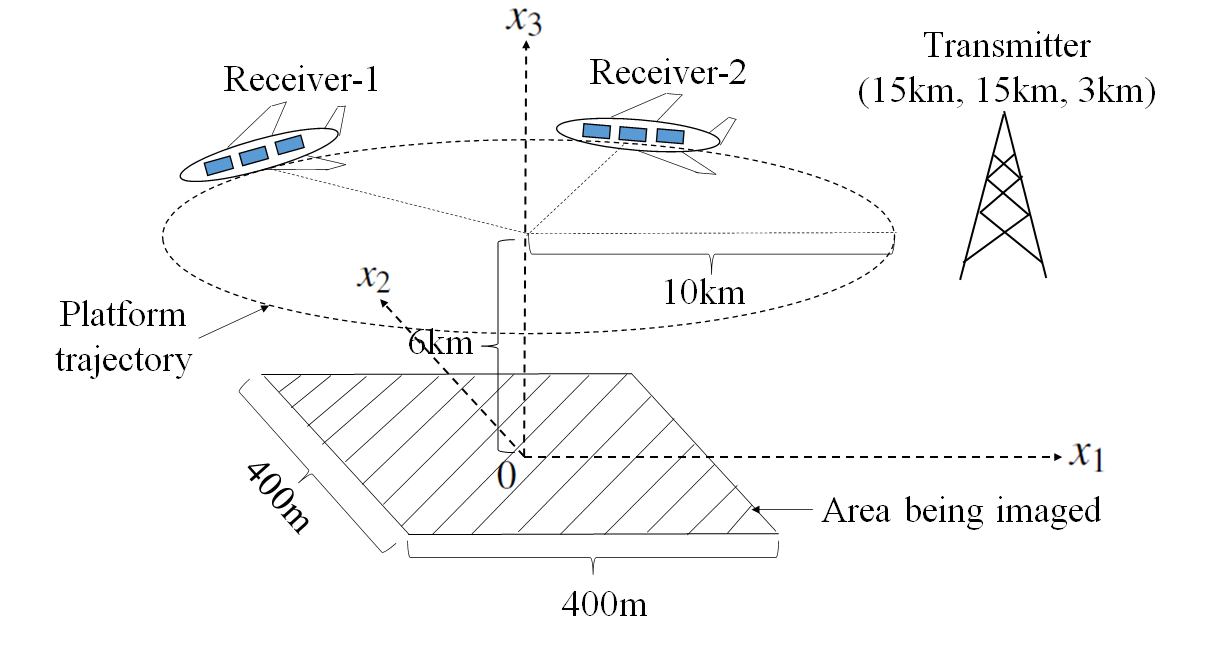}
\caption{Data collection model for passive bistatic SAR.}
\label{fig:passive_SAR_simulations}
\end{figure}
We use two simulated image sets, each with significantly different target characteristics, and generate the corresponding sets of passive bistatic SAR data using MATLAB.
We assume that we are imaging an area of dimension $400$m$\times 400$m, and it is being reconstructed into $40 \times 40$ pixel and $31 \times 31$ pixel images for the first and the second datasets, respectively.
The first dataset includes $5,000$ training and $50$ test samples, and each scene is composed of arbitrarily located random number of point targets.
The second dataset, on the other hand, contains $10,000$ training and $50$ test samples, and each scene contains a single randomly located rectangular target of arbitrary dimensions between $0$m to $10$m.
Let the single stationary transmitter be located at $(15, 15, 3)$km, and we assume that the two receivers are traversing the scene along a circle of radius $10$km and at $6$km height from the ground level with its origin located at the scene center.
These receivers are deviated along their respective trajectories by a $45$ degree angle.
Imaging geometry for this passive bistatic SAR configuration is shown in Fig.~\ref{fig:passive_SAR_simulations}.

\subsection{Network Architecture and Reconstructed Images}
\begin{figure*}
\centering
\includegraphics[width=1.4\columnwidth]{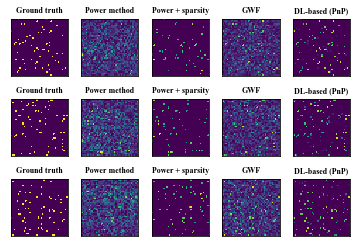}
\caption{For $40 \times 40$ pixel images and $M = 2N$, example reconstruction images using the power method, with and without sparsity prior, GWF algorithms using $150$ updates and our PnP based imaging network with $8$ update stages are shown in the second to fifth column, consecutively. The first column shows the corresponding ground truth images.}
\label{fig:reconstructed_images1}
\end{figure*}

With our end-to-end imaging networks in Fig.~\ref{fig:schematic_diagram} derived using the unrolling technique, we can adopt a supervised training scheme that uses the cross-correlated measurements and the corresponding ground truth images.
Once an optimal set of denoiser parameters are learned, we use these values for imaging from new cross-correlated measurement vectors in the test dataset.
We model the denoisers for both PnP and RED based imaging networks using $16$ layer CNNs with a $3 \times 3$ dimensional convolution filter at each layer.
We use $16$ output channels and $\text{leaky}\_\text{relu}(.)$ activation functions at the hidden layers, and apply $\text{relu}(.)$ function at the output layer.
During implementation, we used the same set of denoiser parameter values for two consecutive layers.
While implementing our imaging networks in Fig.~\ref{fig:schematic_diagram} for the datasets with point targets and rectangular targets, we used $L$ equal to $8$ and $4$, respectively.

\begin{figure*}
\centering
\includegraphics[width=1.4\columnwidth]{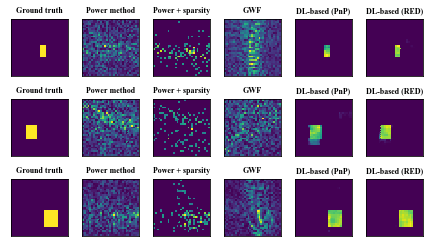}
\caption{For $31 \times 31$ pixel images and using $M = 1.07N$ and $10$dB SNR, example ground truth and reconstructed images using the power method, power method with sparsity prior, GWF algorithm, and our imaging networks based on PnP and RED algorithms using $4$ update stages are shown in the seven columns.}
\label{fig:reconstructed_images_extended}
\end{figure*}

Example reconstruction results for our PnP based network for the two datasets are shown in the last column in Fig.~\ref{fig:reconstructed_images1} and the second last column in Fig.~\ref{fig:reconstructed_images_extended}.
The last column in Fig.~\ref{fig:reconstructed_images_extended} shows the estimated images using our RED based network with the same denoiser architecture and the number of RNN stages as its PnP counterpart.
All the estimated images in Fig.~\ref{fig:reconstructed_images1} are obtained using $80$ slow-time and $40$ fast-time points.
On the other hand, for the second dataset with single rectangular objects, we collected the measurements at $32$ slow-time and $32$ fast-time points, and considered additive Gaussian noise on the correlated measurements with $10$dB SNR.
The second, third and fourth columns in Fig.~\ref{fig:reconstructed_images1} and~\ref{fig:reconstructed_images_extended} display the interferometric inversion results using the power method based spectral estimation, power method augmented by proximal operator under sparsity prior and the GWF algorithm using $150$ updates, respectively.
We observe that for both datasets, our proposed deep denoising prior based imaging algorithm outperforms state-of-the-art methods.
Aside from improved reconstruction quality, we observed improved computation time offered by our imaging network.
For example, the GWF implementation in Fig.~\ref{fig:reconstructed_images_extended} using $150$ updates took $12.5925$s per test samples whereas our PnP based imaging network took only $0.0053$s.

\subsection{Sample Complexity}
\begin{figure*}[!t]
\centering
\subfloat[]{\includegraphics[width=0.65\columnwidth]{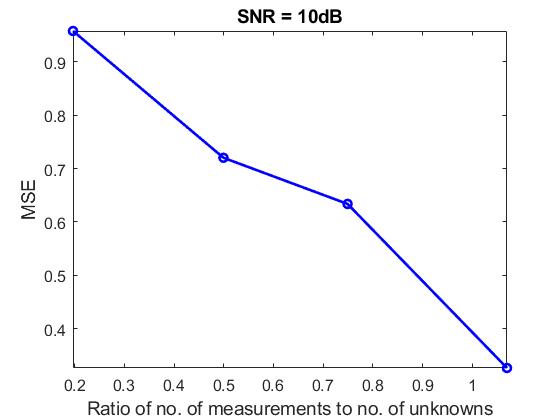}%
\label{fig_mse_vs_MbyN}}
\subfloat[]{\includegraphics[width=0.65\columnwidth]{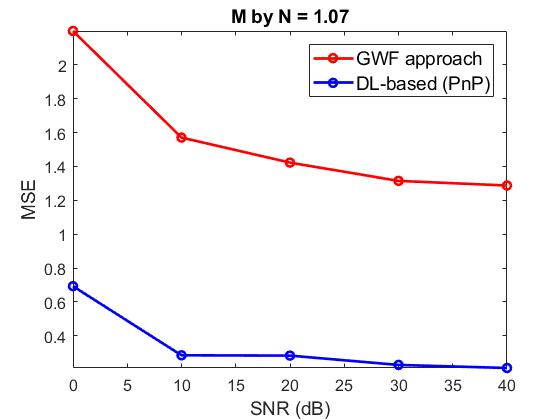}%
\label{fig_mse_vs_snr}}
\caption{Plots of MSE values versus the $\frac{M}{N}$ ratios in (a) and the SNR values in (b) calculated for the test dataset with rectangular targets.}
\label{fig:mse_vs_MbyN_snr}
\end{figure*}
 We perform training and reconstructions for the dataset with rectangular objects by applying the same PnP based imaging network using different numbers of measurements.
The resulting MSE values versus the $\frac{M}{N}$ ratios are displayed in Fig.~\ref{fig_mse_vs_MbyN}.
We observe that for attaining good reconstruction quality using our PnP based network, it is important to have sufficiently large $M$ for a particular image dimension.
However, at each $\frac{M}{N}$ value, reconstruction quality obtained using our approach is significantly better compared to that of the GWF algorithm.
For example, at the $4$ consecutive $M/N$ values in Fig.~\ref{fig_mse_vs_MbyN}, the MSE values obtained using the GWF algorithm are $1.3690$, $1.3641$, $1.3343$ and $1.5701$, respectively.
This empirically shows that our deep denoising and spectral estimation based approach overcomes the strict sample complexity limitation imposed by the GWF algorithm.

\subsection{Effect of SNR}
We varied the additive Gaussian noise levels at the cross-correlated measurements for the dataset with rectangular objects and plotted the corresponding normalized MSE values in Fig.~\ref{fig_mse_vs_snr}.
We observe that as expected, higher SNR values are conducive to better reconstruction quality by our imaging network.
Moreover, compared to the corresponding MSE values attained by the GWF algorithm, our approach shows better robustness to additive noise compared to the state-of-the-art.

\section{Conclusions}
In this paper, we presented two deep denoising prior based interferometric imaging networks whose architectures are rooted in underlying iterative algorithms minimizing either an implicit or an explicit objective function.
Aside from the passive bistatic SAR setting considered in this paper, our approach is also suitable for the interferometric imaging problem for multi-static radar.
Our numerical simulation results empirically demonstrated several benefits offered by our approach over the state-of-the-art including improved accuracy, computation time, sample complexity and noise robustness.
Theoretically determining sufficient conditions for exact recovery and studying the associated restrictions imposed on the physical imaging parameters, as well as relating the minimum required number of updates for attaining specific accuracy levels using our deep denoising prior based algorithms are important open questions to be studied in a future work.

\bibliographystyle{IEEEtran}
\bibliography{bib_file, references}

\end{document}